\documentclass[preprint,proceedings]{rmaa}
\usepackage{paralist}
\usepackage{psfrag,color}
\newcommand{\lppr}{\stackrel{<}{\scriptstyle \sim}}
\newcommand{\lappr}{\raisebox{-0.4ex}{$\lppr $}}
\SetYear{2003}
\SetConfTitle{Compact Binaries in the Galaxy and Beyond}

\title{Modeling of Oxygen-Neon Dominated Accretion Disks in Ultracompact 
       X-ray Binaries: 4U\,1626-67} 
\author{K. Werner,\altaffilmark{1}
T. Nagel,\altaffilmark{1}
S. Dreizler,\altaffilmark{2}
and T. Rauch\altaffilmark{3,1}}

\altaffiltext{1}{IAAT, Universit\"at T\"ubingen, Germany}
\altaffiltext{2}{Universit\"atssternwarte G\"ottingen, Germany}
\altaffiltext{3}{Sternwarte Bamberg, Germany}

\shortauthor{Werner et al.}
\shorttitle{Modeling O-Ne Accretion Disks in Ultracompact Binaries}

\fulladdresses{
\item Stefan Dreizler: Univ.-Sternwarte, D-73083 G\"ottingen, Germany 
  (\email{dreizler@astro.physik.uni-goettingen.de)}.
\item Thorsten Nagel, Thomas Rauch, Klaus Werner, Institut f\"ur Astronomie und
  Astrophysik, Universit\"at T\"ubingen, Sand 1, D-72076 T\"ubingen, Germany 
  (\email{nagel, rauch, werner@astro.uni-tuebingen.de}).
}

\listofauthors{K. Werner, T. Nagel, S. Dreizler, \& T. Rauch}
\indexauthor{Werner, K.}
\indexauthor{Nagel, T.}
\indexauthor{Dreizler, S.}
\indexauthor{Rauch, T.}

\abstract{We report on first results of computing synthetic spectra from
H/He-poor accretion disks in ultracompact LMXBs. We aim at the determination of
the chemical composition of the very low-mass donor star, which is the core of a
former C/O white dwarf. The abundance analysis allows to draw conclusions on
gravitational settling in WDs which is an important process affecting cooling
times and pulsational g-mode periods.}
 
\addkeyword{Accretion disks}
\addkeyword{Binaries: close}
\addkeyword{White dwarfs}
\begin{document}
\maketitle

\begin{figure*}[!t]
  \includegraphics[width=14cm]{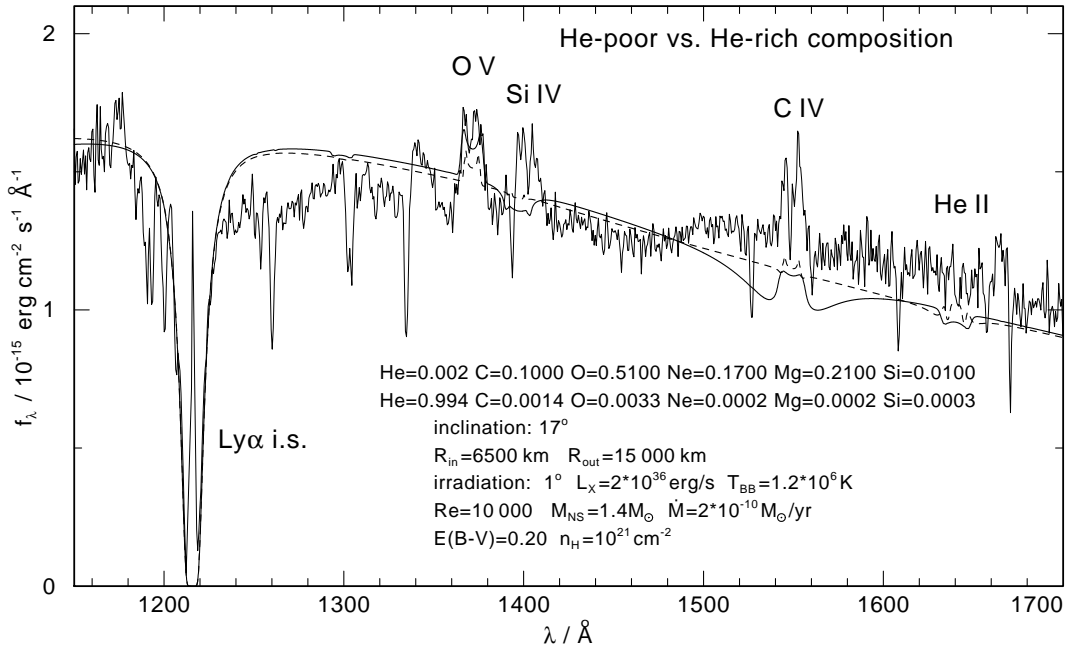}
\caption[]{HST/STIS accretion disk spectrum of 4U\,1626-67 and two model 
spectra, differing by their helium abundance. One is He-poor (0.2\% by mass,
full line), the other is He-rich (99\%, dashed line). The observed lack of
\ion{He}{II}~1640\AA\ obviously is no proof for the He-deficiency of the
disk. Note the broad absorption wings of the \ion{C}{IV} resonance line in the
He-poor model, which are not observed. This points at a stronger irradiation
than assumed.}
\end{figure*}

\section{Properties of 4U\,1626-67}

4U\,1626-67 is a low-mass X-ray binary with an orbital period of 42~min. The
accretor is a 7.7s X-ray pulsar with a magnetic field strength of
$3\cdot10^{12}$\,G (Orlandini \etal\ 1998). The donor is a very low mass
degenerate star with M=0.02--0.08\,M$_\odot$ (Chakrabarty 1998). Its progenitor
is probably a white dwarf with a C/O core. The enhanced neon abundance observed
in the accretion disk is the result of chemical fractionation within the white
dwarf core (Yungelson \etal\ 2002). According to Chakrabarty (1998) the
accretion disk has an inner radius of 6500\,km (corotation radius) and the outer
radius is tidally truncated at 200\,000\,km. The mass-transfer rate amounts to
$2 \cdot 10^{-10}$\,M$_\odot$/yr and the inclination angle is either close to
8$^\circ$ or 33$^\circ$.
Chandra spectroscopy (Schulz \etal\ 2001) suggests that the disk's
chemical composition is O-Ne rich. The observed emission lines from highly
ionized O and Ne are double peaked and probably stem from the Keplerian rotating
accretion disk. Absorption edges in the X-ray spectra point to a C/O WD
donor. HST/STIS observations appear to corroborate the H-He poor chemistry in
the disk. The UV spectra show double peaked emission lines, e.g. from
\ion{C}{IV} and \ion{O}{V} but do not show the \ion{He}{II}~1640\AA\ line (Homer
\etal\ 2002).  A quantitative spectral analysis of the accretion disk
composition would allow the test the idea that the donor is a stripped C/O white
dwarf. We could also determine abundances of heavier elements (Ne, Mg), which
would enable us to quantitatively test theories about gravitational settling of these
elements in WD cores. This process is intensively debated because it
significantly affects  WD cooling times and also g-mode periods in pulsating WDs
(e.g.\ Deloye \& Bildsten 2002).

4U\,1626-67 is not unique. It belongs to a small group of six ultracompact LMXBs
($P_{\rm orb} \lappr\,80\,min$; see e.g.\ Juett \etal\ 2001) with very low mass
($ \lappr 0.1$\,M$_\odot$) H-poor donors. The binary separation is of the order
1 light-s (Earth-Moon distance) and mass-transfer is driven by gravitational
radiation. The optical emission is dominated by their X-ray heated accretion
disk and shows no hydrogen lines. This has been reinforced by recent
spectroscopy  of three group members (Nelemans \etal\ 2003).

\section{Disk modeling}

Modeling is performed with a newly developed non-LTE code (Nagel 2003), that is
based on an advanced stellar atmosphere code (Werner \etal\ 2003). Assuming
that the radial disk structure is that of an $\alpha$-disk (Shakura \& Sunyaev
1973), we model the vertical structure of the disk and the emerging
spectrum as realistic as possible. For that, we divide the disk into a number of
concentric annuli and assume that each annulus radiates as a plane-parallel
slab. We solve consistently the radiation transfer equations and the non-LTE
rate equations for the atomic populations together with the hydrostatic and
energy equations. We can account for full metal line blanketing. This is
important because pressure broadening of spectral lines in the dense LMXB disks
affects the vertical structure by blanketing and backwarming effects. The
kinematic viscosity is parameterized in terms of a Reynolds number (here we set
$Re=10\,000$ which corresponds to $\alpha$=0.01--0.1).

For our model of the hot inner disk regions of 4U\,1626-67, where the UV
spectrum is formed, we assume $M_{\rm NS}$=1.4\,M$_\odot$.  The adopted disk
composition considers the results from X-ray spectroscopy (Schulz \etal 2001). 
We assume a H-free
and strongly He-poor disk with high amounts of O and Ne (see Fig.\,1). For
comparison we computed a He-dominated disk with a composition that represents an
AM\,CVn disk chemistry. We also assume that the disk is irradiated by a central
source with a blackbody spectrum with T=1.2$\cdot 10^6$\,K and a luminosity of
L=2$\cdot 10^{36}$\,erg/s. We present first results of our attempts to fit the
HST/STIS spectrum in Fig.\,1. The spectral lines can be modeled qualitatively
and the main result is, that a He-rich disk composition cannot be ruled out from
this observation. We will perform VLT optical spectroscopy to look for
\ion{He}{II}~4686\AA\ and our next step aims at analyzing the X-ray emission
line spectra.

\acknowledgements

This work was supported by DFG (We 1312/24) and DLR (50\,OR\,0201). We thank Lee
Homer for sending us his reduced HST/STIS spectrum.

\end{document}